\documentclass[journal,twoside,web]{ieeecolor}
\usepackage{tmp_st}
\usepackage[utf8]{inputenc}
\usepackage[margin=0.75in]{geometry}
\usepackage{multirow}
\usepackage{amsmath,amsfonts,amssymb}
\usepackage{graphicx}
\usepackage{cite}
\usepackage{textcomp}
\def\BibTeX{{\rm B\kern-.05em{\sc i\kern-.025em b}\kern-.08em
    T\kern-.1667em\lower.7ex\hbox{E}\kern-.125emX}}
\usepackage{arydshln}

\date{}

\begin{document}
\onecolumn
This work has been submitted to the IEEE for possible publication. Copyright may be transferred without notice, after which this version may no longer be accessible
\twocolumn
\newpage
\title{Assessing the capacity of a \\ denoising diffusion probabilistic model \\ to reproduce spatial context}

\author{Rucha Deshpande, Muzaffer \"{O}zbey, Hua Li, \\ Mark A. Anastasio, \textit{Senior Member, IEEE}, and Frank J. Brooks
\thanks{This work was supported in part by NIH Awards EB031772 (subproject 6366), EB031585 and CA238191. 
Rucha Deshpande is with the Department of Biomedical Engineering, Washington University in St. Louis, St. Louis, MO 63130 USA (email: r.deshpande@wustl.edu) 
Muzaffer \"{O}zbey is with the Department of Electrical and Computer Engineering, University of Illinois at Urbana–Champaign, Urbana, IL 61801 USA (e-mail: mozbey2@illinois.edu).
Hua Li and Mark A. Anastasio are with the Department of Bioengineering, University of Illinois at Urbana–Champaign, Urbana, IL 61801 USA (e-mail: \{huali19,maa\}@illinois.edu).
Frank J. Brooks is with the Center for Label-free Imaging and Multiscale Biophotonics (CLIMB) at the University of Illinois Beckman Institute, Urbana, IL 61801 USA (email: fjb@illinois.edu). \textit{(Corresponding author: Frank J. Brooks.)} The first two authors contributed equally, and the last two authors contributed equally. 
}}
\maketitle

\begin{abstract}
Diffusion models have emerged as a popular family of deep generative models (DGMs). In the literature, it has been claimed that one class of diffusion models---denoising diffusion probabilistic models (DDPMs)---demonstrate superior image synthesis performance as compared to generative adversarial networks (GANs). To date, these claims have been evaluated using either ensemble-based methods designed for natural images, or conventional measures of image quality such as structural similarity. However, there remains an important need to understand the extent to which DDPMs can reliably learn medical imaging domain-relevant information, which is referred to as `spatial context' in this work. To address this, a systematic assessment of the ability of DDPMs to learn spatial context relevant to medical imaging applications is reported for the first time. A key aspect of the studies is the use of stochastic context models (SCMs) to produce training data. In this way, the ability of the DDPMs to reliably reproduce spatial context can be quantitatively assessed by use of post-hoc image analyses. Error-rates in DDPM-generated ensembles are reported, and compared to those corresponding to a modern GAN. The studies reveal new and important insights regarding the capacity of DDPMs to learn spatial context. Notably, the results demonstrate that DDPMs hold significant capacity for generating contextually correct images that are `interpolated' between training samples, which may benefit data-augmentation tasks in ways that GANs cannot.
\end{abstract}

\begin{IEEEkeywords}
denoising diffusion probabilistic models, deep generative model evaluation, medical image synthesis, stochastic context models, stochastic object model \vspace{-0.25cm}
\end{IEEEkeywords}

\section{Introduction}

Deep generative models (DGMs) have shown tremendous potential for advancing medical imaging research \cite{kazeminia2020gans,shamsolmoali2021image,kazerouni2023diffusion,khader2023denoising,dorjsembe2022three}. Significant advancements in DGMs have been achieved in the last few years \cite{bond2021deep,song2019generative,song2020score,song2022solving,karras2020analyzing}. Recently, a novel paradigm based on diffusion generative modeling \cite{sohl2015deep} has been actively developed and explored in medical image research \cite{kazerouni2023diffusion,khader2023denoising,dorjsembe2022three,CardosoLatentDiffusion}. Three related formulations appear in the diffusion modeling literature: denoising diffusion probabilistic models (DDPMs) \cite{ho2020denoising,nichol2021improved,dhariwal2021diffusion}, score-based generative models (SGMs) \cite{song2019generative}, and stochastic differential equations (SDEs) \cite{song2020score}. The first formulation, DDPM \cite{ho2020denoising,nichol2021improved,dhariwal2021diffusion}, is chosen as the focus of this work. DDPMs are designed to estimate the probability density function of the target data distribution by learning to translate an initial noise distribution to the target data distribution over multiple intermediate steps that are modeled as probabilistic transitions. Furthermore, they have been employed for medical imaging applications ranging from medical image synthesis \cite{kazerouni2023diffusion,muller2022diffusion,khader2023denoising,dorjsembe2022three,CardosoLatentDiffusion} to image reconstruction \cite{jalaln2021nips,chung2022media,song2022solving,liu2022dolce}. 

It has been claimed that DDPM demonstrates superior performance in medical image synthesis as compared to generative adversarial networks (GANs) \cite{muller2022diffusion,dorjsembe2022three,CardosoLatentDiffusion,huy2023denoising,iskandar2023towards}. However, this claim was based on comparisons that primarily relied on ensemble-based evaluation measures employed in natural image feature spaces such as Fr\'echet Inception Distance (FID)\cite{heusel2017gans}, precision and recall\cite{salimans2016improved}, or conventional image quality measures such as the structural similarity index measure (SSIM)\cite{wang2004image}. The relevance of these evaluation measures to medical image assessment remains unknown. Similarly, other evaluation approaches developed primarily for natural images \cite{borji2022pros,borji2019pros,akbar2023beware,akbar2023brain} also may not be sufficient for assessing downstream utility in medical imaging tasks. On the other hand, visual evaluation of medical images \cite{chuquicusma2018fool, park2021realistic} is challenged by the need for domain expertise. Thus, although innovations in DGMs have been translated rapidly from computer vision to medical imaging research, the strategies employed to evaluate DGMs are rarely tailored for medical imaging applications \cite{kazeminia2020gans,shamsolmoali2021image,kazerouni2023diffusion} and there remains an urgent need for an objective evaluation of diffusion models to determine their applicability for biomedical imaging.

In this work, the suitability of a DDPM for medical imaging applications is tested via its capacity to reproduce prescribed spatial context, represented via certain per-image statistics. Our evaluation approach is similar to some previous works \cite{deshpande2021method, kelkar2023assessing}. Here, ``context'' is defined as domain-relevant information. Context cannot necessarily be learned from just one image, but can be explicitly encoded, or implicitly emergent. Examples are: the prevalence of image classes (i.e., medical conditions) in an ensemble, the size and shape of specific organs, anatomical constraints such as the number of ribs or \emph{relative} size/location of organs, or the organ-specific appearance of image texture. Contextual errors have previously been reported in GAN-generated ensembles \cite{deshpande2021method, dumont2021overcoming, muller2022diffusion, kelkar2023assessing}, but their occurrence in DDPM-generated ensembles has not been studied. Assessing the reproducibility of context \cite{oliva2007role, emrith2010measuring, doersch2014context} provides one way of gaining insights into the performance of DDPM-generated ensembles for medical imaging applications. 

Specifically, a DDPM was first tested for its capacity to reproduce \emph{explicit} context via a previously established test bed of stochastic context models (SCMs) \cite{deshpande2021method}. The SCMs represent attributes relevant to medical imaging, in a readily interpretable manner and without anatomical constraints. Next, the DDPM was tested for its capacity to reproduce \emph{implicit} context via an adapted version of a previously published stochastic object model (SOM) that describes anatomical constraints \cite{badano2018evaluation}. Per-image, contextual errors in DDPM-generated ensembles were then quantified to provide a measure of the potential suitability of DDPM to medical imaging tasks involving similar contextual attributes. This evaluation approach can be employed to gain insights into other DGM formulations as well. 

\section{Background}
\subsection{Denoising diffusion probabilistic models (DDPM)}

In the DDPM framework \cite{dhariwal2021diffusion}, a small quantity of Gaussian noise is gradually injected into an input image (sampled from a real data distribution) $x_0 \sim q(\boldsymbol{X}_0) $ over $t$ time steps to eventually obtain the degraded image $x_t$. Over a sufficiently large number of time steps $T$, a sample $x_T$ from a Gaussian distribution can be produced. The forward diffusion process is formulated as a Markov chain where the  $x_t$ and $x_{t-1}$ are related by the transition rule defined as:
\begin{eqnarray}
&& \boldsymbol{x}_{t}=\sqrt{1-\beta_{t}}\boldsymbol{x}_{t-1}+\sqrt{\beta_{t}}\boldsymbol{\epsilon},\quad{}\boldsymbol{\epsilon}\sim \mathcal{N}\left( \boldsymbol{0},\boldsymbol{I} \right) \\
&& q\left( \boldsymbol{x}_{t}|\boldsymbol{x}_{t-1} \right)=\mathcal{N}\left( \boldsymbol{x}_{t}; \sqrt{1-\beta_{t}}\boldsymbol{x}_{t-1},\beta_{t}\boldsymbol{I} \right).
\end{eqnarray}
Here, $\mathcal{N\left( \boldsymbol{0},\boldsymbol{I} \right)}$ is a Gaussian distribution with zero mean and identity covariance $I$, $\beta_t$ is noise variance, and $\boldsymbol{\epsilon}$ is additional noise. The reverse diffusion process that maps $x_T$ to $x_0$ is also formulated as a Markov chain, where each step represents an incremental denoising of the data. The reverse transition probability between $x_{t-1}$ and $x_t$ can be represented by a Gaussian distribution for a large $T$ and small $\beta_t$:
\begin{equation} \label{eq:backward}
	q(\boldsymbol{x}_{t-1}|\boldsymbol{x}_{t}) := \mathcal{N}(\boldsymbol{x}_{t-1}; \boldsymbol{\mu}(\boldsymbol{x}_t, t), \mathbf{\boldsymbol{\Sigma}}(\boldsymbol{x}_t, t)).
\end{equation}
Within each reverse diffusion step, the same neural network with time embedding is employed to approximate the reverse mapping by predicting the mean ($\boldsymbol{\mu}$) and covariance ($\boldsymbol{\Sigma}$). The variational lower bound ($L_{\textrm{vb}}$) was employed in the loss function to minimize the negative log-likelihood:
\begin{equation} \label{eq:lvlb}
    L_{vb} = \mathbb{E}_{q(\boldsymbol{x}_{0:T})} \left[ \, \textrm{log} \, \frac{q(\boldsymbol{x}_{1:T} | \boldsymbol{x}_0)}{p_{\theta}(\boldsymbol{x}_{0:T})} \right] \ge - \mathbb{E}_{q(\boldsymbol{x}_{0})} \left[ \, \textrm{log} \, p_{\boldsymbol{\theta}}(\boldsymbol{x}_0) \right], 
\end{equation}
where $ p_{\boldsymbol{\theta}}$ is the network parameterization for the approximated reverse diffusion process,  $\boldsymbol{\theta}$ represents the network parameters, and $\mathbb{E}_q$ represents expectation over $q$. The collection of data samples between time steps 0 and $T$ is represented by $\boldsymbol{x}_{0:T}$, while the image samples between time steps 1 and $T$ conditioned on the sample at time step 0 are represented by $\boldsymbol{x}_{1:T} | \boldsymbol{x}_0$. 
The bound can be reformulated as\cite{dhariwal2021diffusion}:
\begin{eqnarray}
\label{eq:Lall}
\hspace{-3.2mm}
L_{vb} &=& \textrm{log} \, p_{\boldsymbol{\theta}} (\boldsymbol{x}_0 | \boldsymbol{x}_1) \nonumber \\ 
&&- \sum_{t=1}^{T} \textrm{KL}(q(\boldsymbol{x}_{t-1} | \boldsymbol{x}_t , \boldsymbol{x}_0) \, || \, p_{\boldsymbol{\theta}}(\boldsymbol{x}_{t-1}|\boldsymbol{x}_t)), 
\end{eqnarray}
where KL denotes Kullback-Leibler divergence, and $\textrm{KL}(q(\boldsymbol{x}_T | \boldsymbol{x}_0) \, || \, p(\boldsymbol{x}_T))$ is omitted as it does not depend on $\boldsymbol{\theta}$. With the reparameterization of the second term in Eq. (\ref{eq:Lall}), the employed network can predict $\boldsymbol{\epsilon_t}$. In Ref. \cite{dhariwal2021diffusion}, the mean and covariance were learned jointly:
\begin{eqnarray}
\label{eq:KL}
\hspace{-3.2mm}
 \textrm{KL}(q(\boldsymbol{x}_{t-1} | \boldsymbol{x}_t , \boldsymbol{x}_0) \, || \, p_{\boldsymbol{\theta}}(\boldsymbol{x}_{t-1}|\boldsymbol{x}_t))  \nonumber \\
= \frac{1}{2||\boldsymbol{\Sigma}_{\boldsymbol{\theta}} || _2^2} ||\boldsymbol{\mu}(\boldsymbol{x}_t, \boldsymbol{x}_0) - \boldsymbol{\mu_\theta}(\boldsymbol{x}_t, t)  ||^2  \nonumber \\
= \frac{(1-\alpha_t)^2}{2\alpha_t(1-\Bar{\alpha_t})||\boldsymbol{\Sigma}_{\boldsymbol{\theta}} || _2^2} ||\boldsymbol{\epsilon_t} - \boldsymbol{\epsilon_\theta}(\boldsymbol{x}_t, t)  ||^2.
\end{eqnarray}

\subsection{Description of the stochastic context models (SCMs)}
The following subsections provide brief descriptions of the three SCMs employed in this work. Each SCM encodes a different set of contextual constraints (see Table~\ref{table_scms}), and is either single-, or multi-class. Further details can be found in a previous work \cite{deshpande2021method}, and the data are publicly available \cite{DVN/HHF4AF_2021}. A realization from any SCM is a 256$\times$256, grayscale image. The single-class ensemble contained 131072 images, and the multi-class ensembles contained 65536 images per class.

\subsubsection{Alphabet SCM (A-SCM)}
The A-SCM represents prescribed contextual constraints of prevalence and relative positions of features. This SCM is analogous to anatomical scenarios that exhibit constraints in per-image prevalence and relative positions of structures. A realization of the A-SCM corresponds to a regular grid of 32$\times$32-pixel tiles $t$; each tile represents a letter in the alphabet $\mathbb{A} = \{H, K, L, V, W, X, Y, Z\}$. The per-image frequency of occurrence of all letters was prescribed such that each realization $I$ consisted of the exact set of letters: $\mathbb{B} = \{24 \times H, 2 \times K, 16 \times L, 1 \times V, 1 \times W, 8 \times X, 8 \times Y, 4 \times Z\}$. That is, a realization can be written as:

 \begin{equation}
 \label{alphabet_main}
     I = \{t_{r,c} : \bigcup\limits_{r,c}f(t_{r,c})=\mathbb{B}\},
 \end{equation}
where $f(t)$ represents a template-matching operation that maps an input tile to a letter, and $r, c$ respectively indicate the row and column indices within the grid. The placement of specific letters in a realization \textit{I} was not constrained, thus creating diverse realizations. However, contextual rules of per-image letter as well as letter-pair prevalence were always upheld in each realization. In addition to the prescribed single letter prevalence, rules of conditional letter-pair prevalence were also prescribed:

\begin{equation}
\begin{aligned}
    p(f(t_{r,c+1}) = Y | f(t_{r,c}) = X) = 1,\\
    p(f(t_{r,c}) = Z | f(t_{r+1,c}) \in \{V, W, K\}) = 1.
\end{aligned}
\end{equation}
Each realization included four ordered letter pairs: X-Y (horizontal), Z-K, Z-V, and Z-W (vertical). The respective per-image prevalences of these letter-pairs were 8, 2, 1, and 1. Post-hoc classification of letters in DGM-generated ensemble was performed via template matching.

\subsubsection{Voronoi SCM (V-SCM)}
The four-class V-SCM represents joint contextual constraints in prevalence, intensity and texture. This SCM is logically analogous to microscopy images that contain distinct cell types representative of pathology based on their intensity, texture and prevalence. Specifically, each realization $I$ can be represented as a set $V$, which consists of Voronoi regions $v_i$ and edges $e_i$. Here, $i = \{1, 2, ..., c\}$, where $c \in \{16, 32, 48, 64\}$ denotes the class of image as well as the number of Voronoi regions in an image. The placement of region centers within $I$ was spatially random, introducing positional variance. By setting edges $e$ to an intensity level of 0, the recoverability of Voronoi regions was enhanced. All pixels in a region $v_i$ were allocated a constant grayscale intensity $g$ chosen from a set of 128 predefined values between 1 and 254. Furthermore, the grayscale intensity of a region was perfectly rank-correlated with its area: $\rho(\mathrm{area}(v_i), g) = 1$, where $\rho$ is the Spearman rank-order correlation coefficient. Post-hoc processing of generated realizations to extract individual Voronoi regions involved edge detection via skeletonization and Sauvola thresholding \cite{sauvola2000adaptive}. The error in region detection was not large enough to affect the inferences in our experiments.

\subsubsection{Flag SCM (F-SCM)}
The eight-class F-SCM represents joint contextual constraints in prevalence, intensity, texture, and absolute position; this SCM is logically analogous to the joint constraints specific to organs. Each image $I$, in class $c$, was divided into a 16$\times$16 grid of tiles, where each tile represented either the foreground ($f_k$) or the background ($b_k$), and $k$ denoted the position of the tile within the grid. Thus, each image, irrespective of class, consisted of exactly the same number of foreground and background tiles $I_c = \{80\times f_k, 176\times b_k\}$--- this removed the zero-order variance in the number of foreground pixels across classes. A realization $I$ was modeled as:
\begin{equation}
I_c = \sum_k(a_{kc}f_k + (1-a_{kc})b_k),
\end{equation}

where $\mathcal{A}\in\{0,1\}^{K\times C}$ is a binary matrix signifying background (0) or foreground (1) for all $K$ tile indices in $C$ classes. As a result, the image class is one of eight different foreground arrangements, and $A$ denotes the pre-defined, class-specific foreground patterns. Intensity distributions for $f_k$ and $b_k$ were prescribed as: 
\begin{equation}
\begin{aligned}
    f_k\sim 152\:X + 96,\; \mathrm{where}\; X \sim \mathrm{Beta}(\alpha=4,\beta=2), \\
    b_k\sim 192\:X + 8,\; \mathrm{where}\; X \sim \mathrm{Beta}(\alpha=2,\beta=4). 
\end{aligned}
\end{equation}

The respective variates were placed randomly in the foreground ($f_k$) and background ($b_k$). The presence of multiple constraints not only enabled the testing of the reproducibility of joint constraints at once, but also enabled the testing of preferential learning of any of these constraints. Additionally, a set of 24 tile-location indices ($k$) were forbidden as foreground in any class, this enabled investigation of the nature of interpolation between classes. Comparison of tile-wise intensity means to determine foreground followed by mean absolute error computation against the expected foreground pattern yielded class identity for the DGM-generated ensemble.

\subsection{Description of the adapted VICTRE stochastic object model (VT-SOM)}
The anatomically realistic stochastic model: VT-SOM, is an adapted version\cite{aapm} of a four-class SOM \cite{badano2018evaluation} that describes the anatomy of the human female breast. A brief description of the adaptation procedure follows. First, 8000 realizations of 3D voxelized maps of human numerical breast phantoms (NBPs) were generated and all tissues were allocated linear attenuation co-efficients corresponding to x-rays at energy 30 keV. Fifteen coronal slices were extracted from approximately the central third of a NBP volume and the resulting images were then downsampled to 512$\times$512 while retaining the ligament skeleton. Finally, the tissue types were allocated values in the 8-bit grayscale range largely consistent with their linear attenuation co-efficients. The VT-SOM comprised four breast types – fatty, scattered, heterogeneous and dense – based on the Breast Imaging Reporting and Data System (BI-RADS) \cite{liberman2002breast} classification of breast density. The resulting ensemble of 108,000 slices from the VT-SOM was employed to train DGMs. The VT-SOM has a class prevalence (as determined by the relative proportion of fatty to glandular tissue) in the ratio 1:4:4:1 for the fatty, scattered, heterogeneous and dense breast types. This ratio is based on population prevalence \cite{liberman2002breast}.  


\begin{table}[h!tb]

\caption{Overview of all stochastic context and object models in terms of the \emph{per-image} contextual constraints explicitly prescribed in the model.}
\begin{center}
\begin{tabular}{ccccc}
    \hline
    Constraints  & A-SCM & V-SCM & F-SCM & VT-SOM \\
    \hline
    Prevalence & $\checkmark$ & $\checkmark$ & $\checkmark$ & $\checkmark$ \\ 
    Intensity & $\times$ & $\checkmark$ & $\checkmark$ & $\checkmark$ \\
    Texture & $\times$ & $\checkmark$ & $\checkmark$ & $\checkmark$ \\
    Position & $\checkmark$ & $\times$ & $\checkmark$ & $\checkmark$ \\
    Anatomy & $\times$ & $\times$ & $\times$ & $\checkmark$ \\
    Multi-class & $\times$ & $\checkmark$ & $\checkmark$ & $\checkmark$ \\
    \hline
\end{tabular}
\label{table_scms}
\end{center}
\end{table}

\begin{figure}[hbt]
\centering
\includegraphics[width=0.9\linewidth]{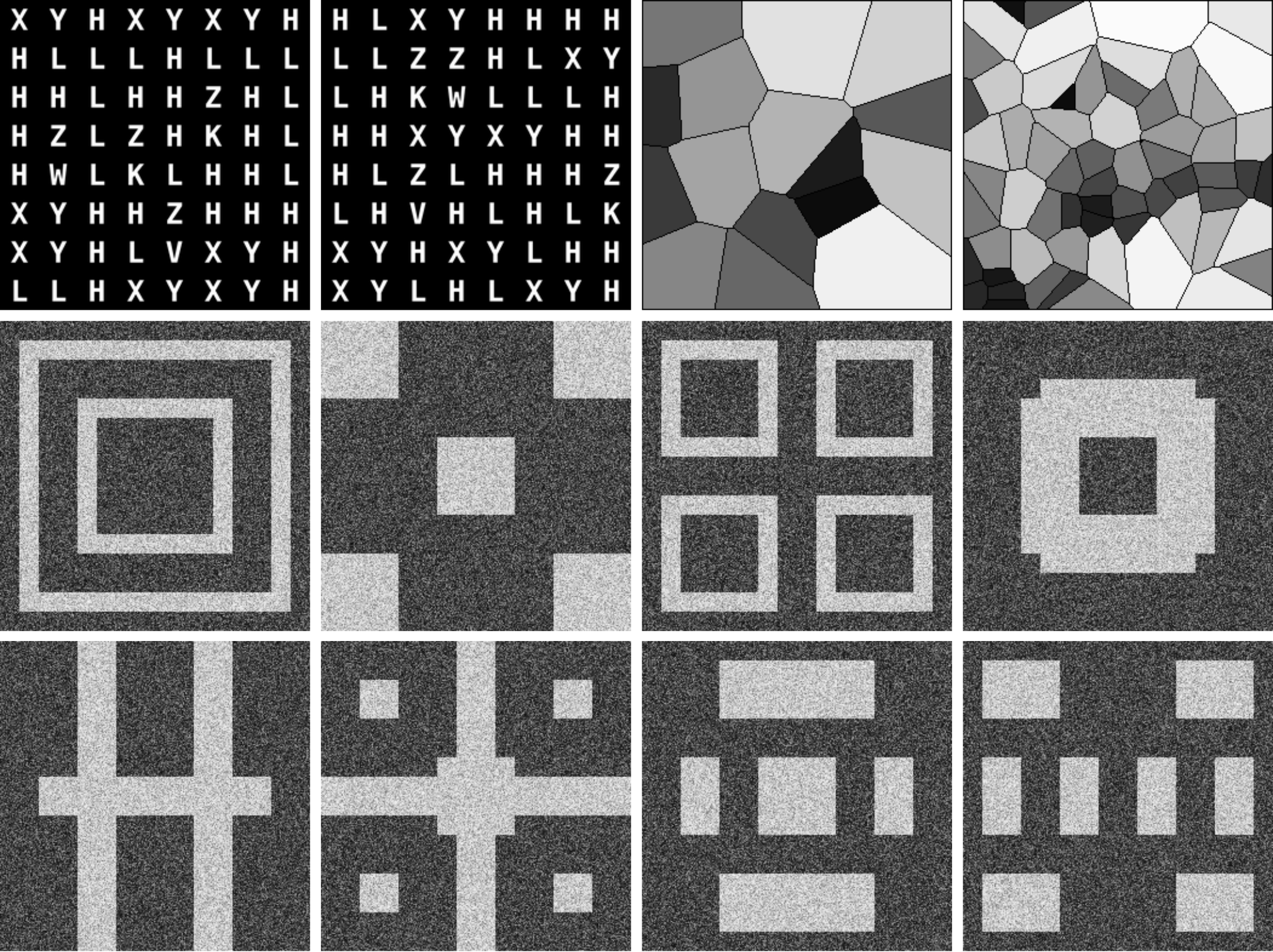} 

\caption{Sample realizations from all three SCMs are shown. Top row: Two realizations each from the single-class A-SCM (left) and the four-class V-SCM (right). Realizations from the V-SCM represent classes 16 and 64 respectively. Rows 2 and 3: A realization from each of the eight classes in the F-SCM.}
\label{SCM_real_ones}
\end{figure}

\begin{figure}[hbt]
\centering
\includegraphics[width=0.9\linewidth]{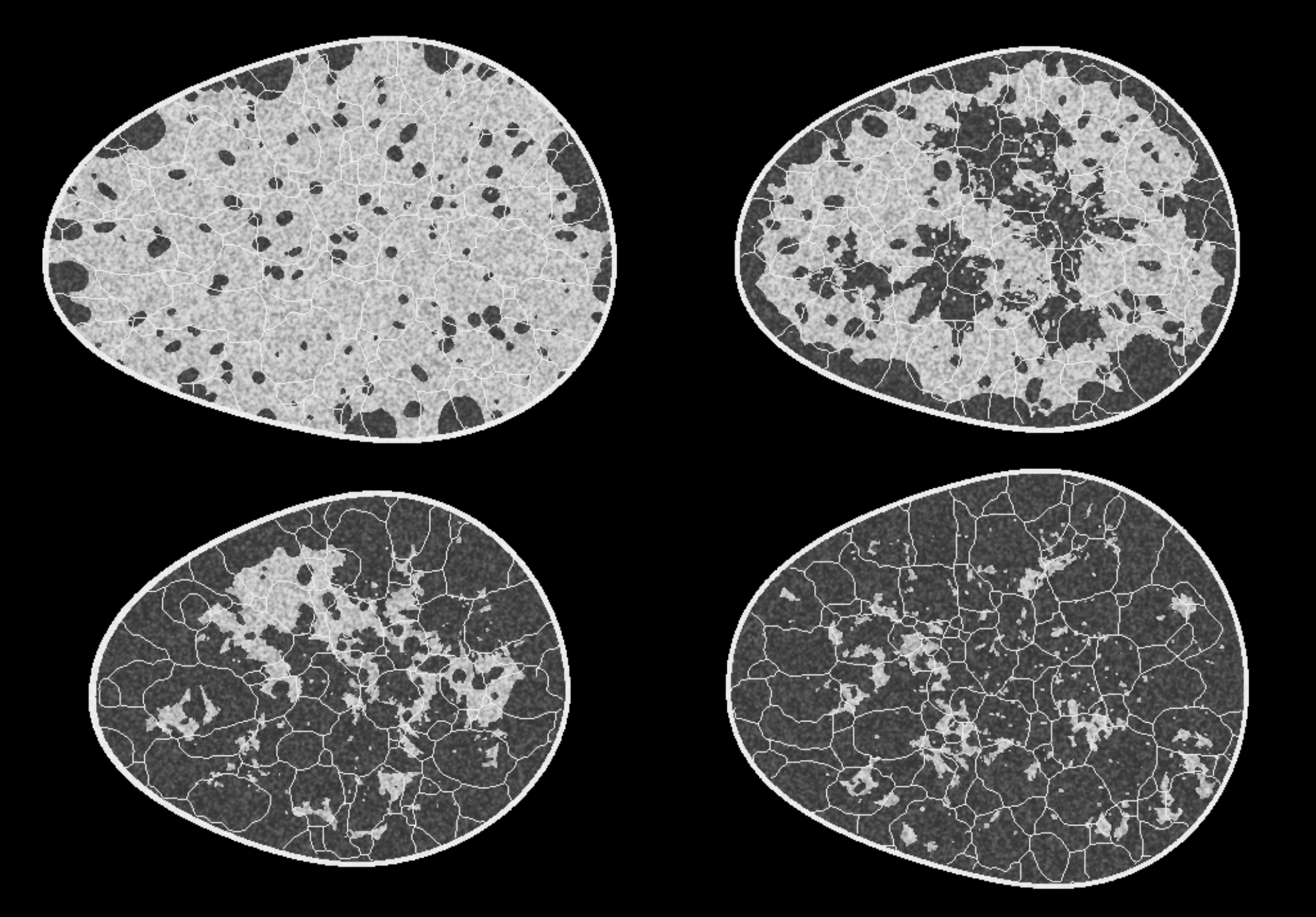} 

\caption{Sample images from each of the four classes in the VT-SOM. Sample realizations from (top row: L to R) dense, heterogeneous, (bottom row: L to R) scattered and fatty breast types are shown.}
\label{victre_real_ones}
\end{figure}

\section{Methods}
\subsection{Network trainings}
A popular GAN architecture, StyleGAN2 (SG2), \cite{karras2020analyzing}, and Denoising Diffusion Probabilistic Model (DDPM)\cite{dhariwal2021diffusion}, two of the most widely utilized DGMs, were employed in this work. Because the recommended default hyperparameters have been optimized for visual quality and Frechet Inception Distance (FID) scores, the defaults were used for all trainings unless specified otherwise. All SG2 models were trained until 25M images were seen by the discriminator. For the SG2 model trained on the VT-SOM, the latent space size was decreased to 256 to improve training stability. All DDPM models were trained with the recommended architectural improvements \cite{dhariwal2021diffusion} and until 10M images were accessed--- although this number was chosen for computational convenience, it was observed to be sufficient for achieving reasonable performance. In case of SG2 trained on the VT-SOM, the model with the least FID was chosen for further analysis; in all other cases, the last model was chosen. A four-class classifier with a VGG-16 backbone \cite{simonyan2014very} was also trained on a distinct ensemble obtained from the VT-SOM to predict the classes in the training dataset. The training and validation datasets consisted of 5000, and 1500 images per class, respectively. The classifier was trained for 400 epochs and the model corresponding to the least validation loss was chosen for inference. The per-class misclassification rates were observed to be 0\%, 0.07\%, 0.37\%, 0.87\% in a test set of 3000 realizations for each of the four classes.

\subsection{Methods: Evaluation framework}
Evaluations for studies involving the SCMs were performed after appropriate post-hoc analysis of the DGM-generated images, as described in a previous work\cite{deshpande2021method}. For the VT-SOM study, first, individual tissues were segmented via global thresholding based on the prescribed tissue-specific intensity distributions. Next, the following feature types were extracted from all training and generated images: skeleton statistics \cite{nunez2018new}, morphological features \cite{van2014scikit}, texture features \cite{haralick1973textural}, and the ratio of fatty tissue to glandular tissue (F/G ratio)\cite{liberman2002breast}. Principal component analysis was performed on these features and the top 10 principal components were retained. The cosine similarity was computed for 10,000, 10-dimensional, randomly selected training-training pairs; this produces an estimated distribution of the similarity among realizations. A second distribution was computed for the same number of randomly selected training-generated pairs. The difference between the two distributions was summarized via the Kolmogorov-Smirnov (KS) test statistic \cite{chakravarti1967handbook}. For class-wise analysis, class coverage and density \cite{naeem2020reliable} were computed in the top two principal component space, while class prevalence was obtained by employing the classifier described in the previous subsection. 

\section{Results}

\begin{figure}[hbt]
\centering
\includegraphics[width=0.95\linewidth]{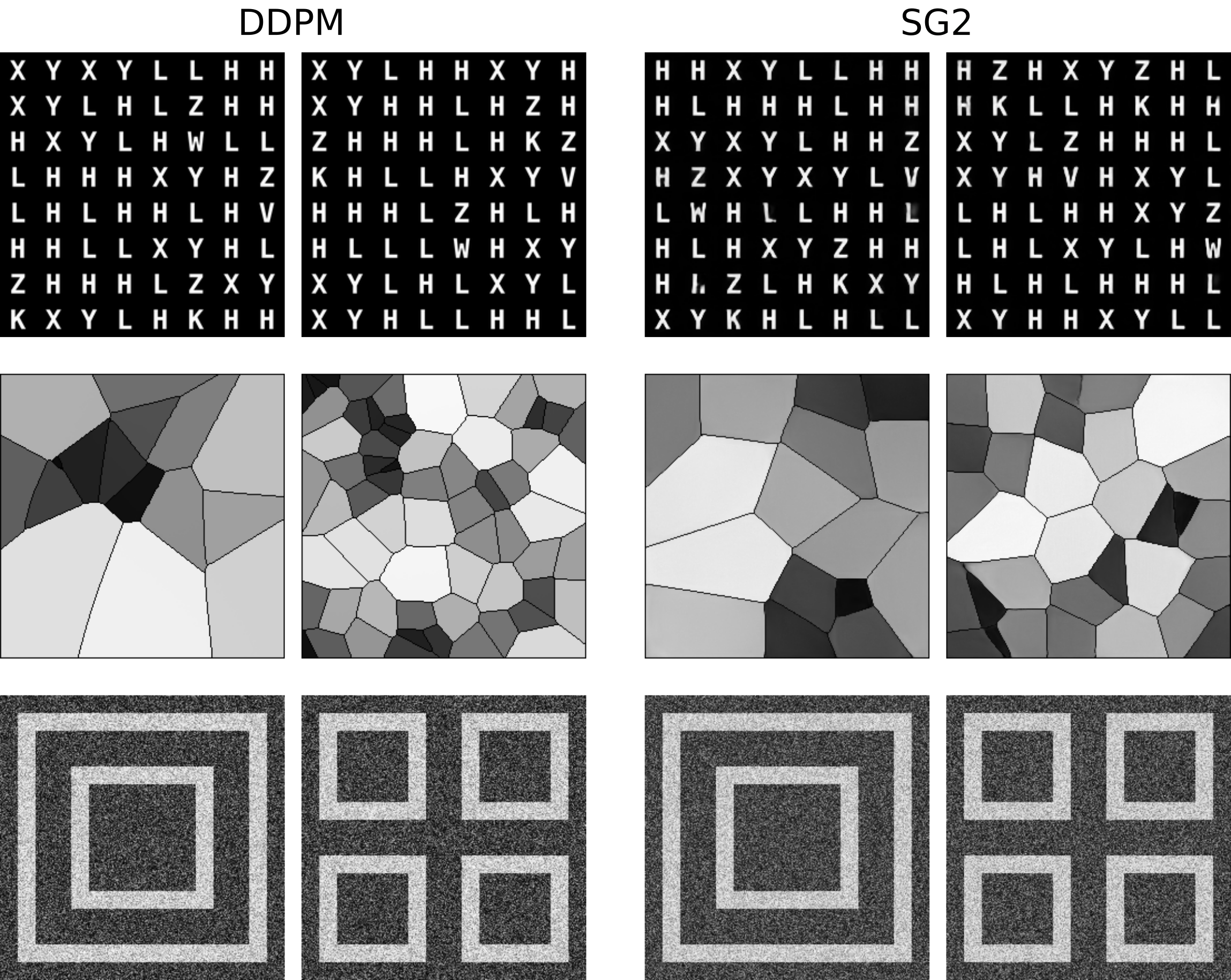} 

\caption{Visually high quality generated samples from DDPM (columns 1 and 2) and SG2 (columns 3 and 4). The rows correspond to generated samples from A-SCM, V-SCM, and F-SCM respectively.}
\label{SCM_good_ones}
\end{figure}

It is noted that the performance of DDPM and SG2 may vary with the choice of training hyperparameters. The performance reported in this work is only representative of typically trained models and may not indicate the best performance possible for either DGM. Sample realizations generated from both DGMs and all stochastic models are shown in Fig.~\ref{SCM_good_ones}. High visual quality was observed in generated images from both DDPM; this is also reflected in the low FID values for all DDPM ensembles. The FID-10k values for A-SCM, V-SCM, F-SCM, and VT-SOM respectively, were: 0.1, 1.5, 5.7, 5.2 for DDPM-generated ensembles, and 6.7, 25.1, 25.7, 9.8 for SG2-generated ensembles. 

\subsection{Results from the A-SCM}

Only realizations within which all letters were visually recognizable were included for further analysis. Recognition was automated via a pattern match filter \cite{deshpande2021method}; the rejection threshold was set to correspond with unambiguous, visually sharp letters as observed in generated images. All DDPM realizations exhibited only recognizable letters, but only 59\% of the SG2 realizations were completely recognizable. Ten-thousand, well-formed realizations from each model were selected randomly for further analysis.
Single letter prevalence was assessed via a chi-squared goodness-of-fit test with the critical value set to 95\%. About 99\% of all DGMs realizations were acceptable, however, only 8 SG2 realizations exhibited perfect letter prevalence; this is in stark constrast to the 9852 DDPM realizations exhibiting perfect prevalence.

\begin{figure}[h!bt]
\centering
\includegraphics[width=0.9\linewidth]{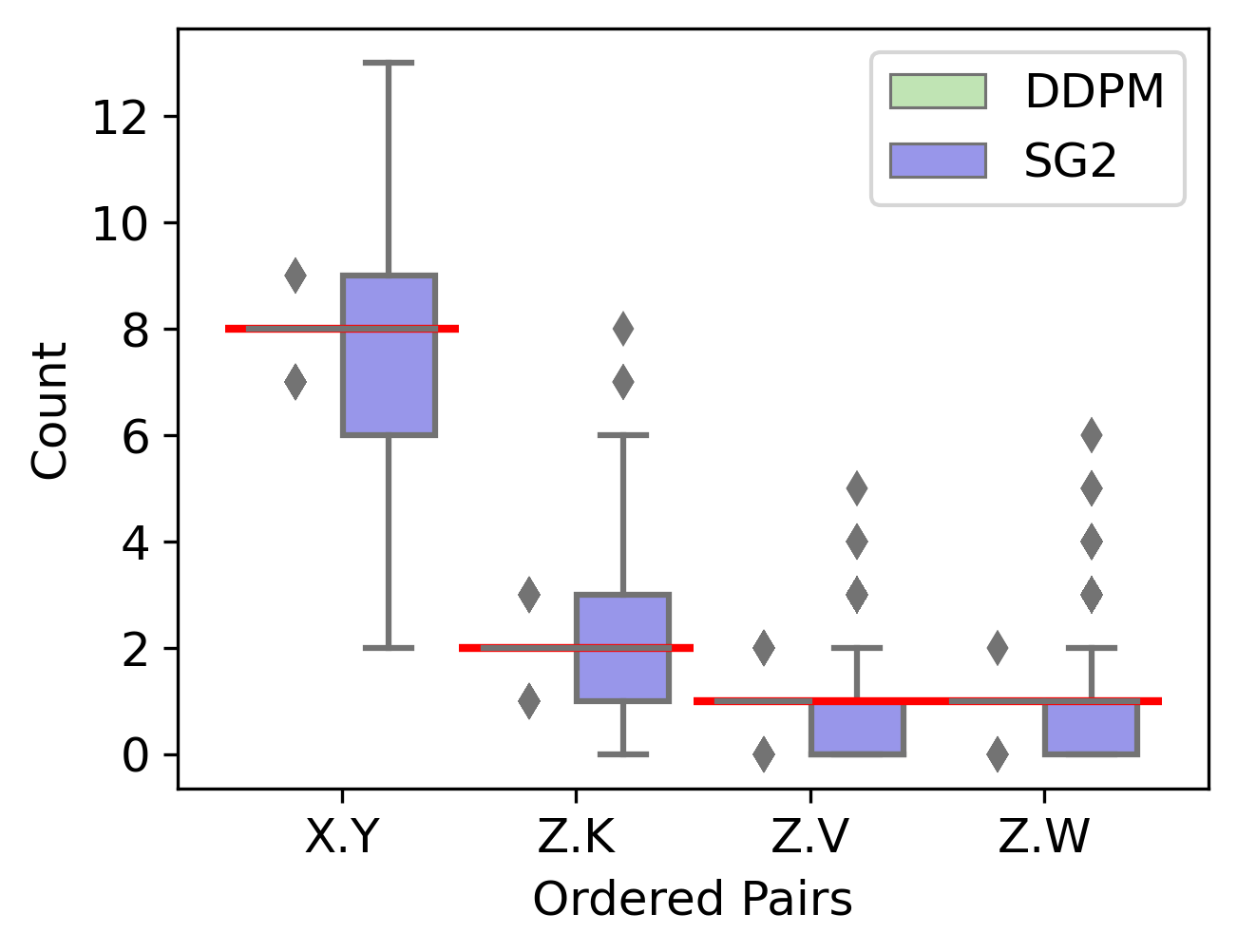} 
\caption{Results from A-SCM. Not only does DDPM clearly outperform SG2, but also achieves near perfect per-image prevalence of the letter-pairs representing contextual constraints. Correct prevalence is marked in red.}
\label{letter_prevalence}
\end{figure}

\begin{figure}[h!bt]
\centering
\includegraphics[width=0.95\linewidth]{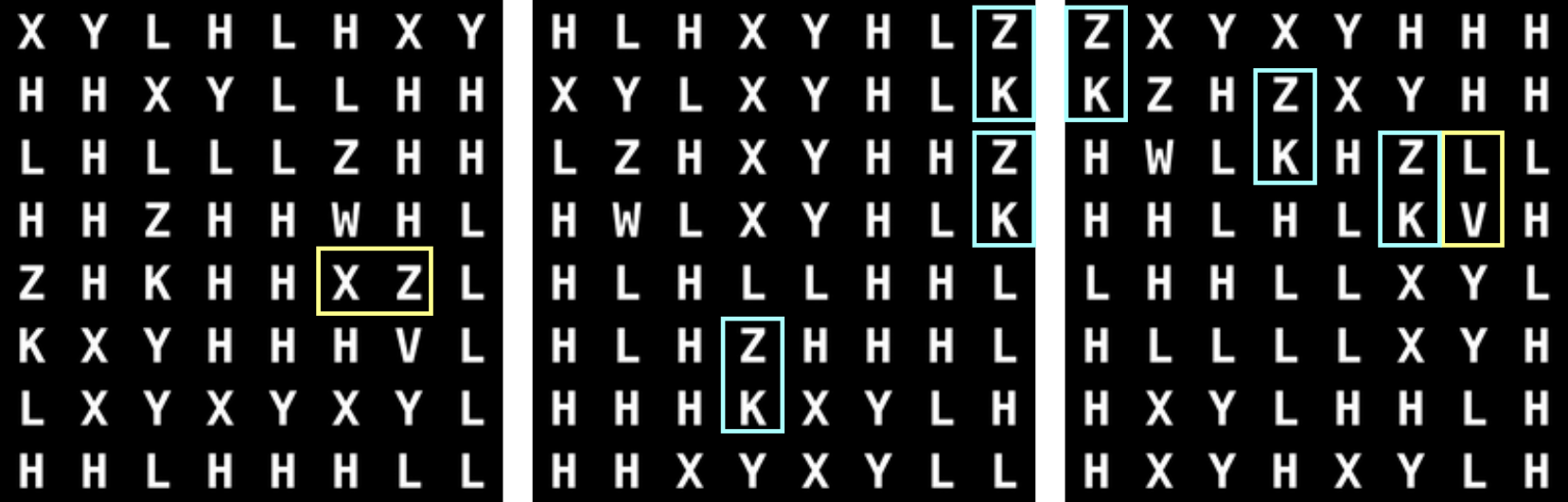} 

\caption{Contextual errors were observed in some DDPM-generated realizations from A-SCM. These manifested as incorrect pairings of letters (yellow) or incorrect per-image prevalence of letter-pairs (blue). In the training data, the letter-pairs X-Y, and Z-V were always in order, and the letter-pair Z-K occurred exactly twice in each image.}
\label{SCM-A-bad_ones}
\end{figure}

Results from the reproducibility of feature-pair prevalences (see Fig.~\ref{letter_prevalence}) strengthen this finding. All four letter-pairs prescribed in the training dataset were almost perfectly replicated throughout the DDPM-generated ensemble, but most SG2-realizations exhibited substantially incorrect letter-pairs prevalence. Some errors in DDPM realizations are shown in Fig.~\ref{SCM-A-bad_ones}. These examples demonstrate that DDPM occasionally creates new pairings, or displays pairings too frequently, even when realizations are otherwise excellent. Thus, by analogy, an ensemble of biomedical images could appear perfect via spot-checks, and pass traditional tests of distribution similarity, but still include images that are anatomically nonsensical.

\subsection{Results from the V-SCM}
DGM-generated images corresponding to the V-SCM were tested for (i) \emph{explicit} contextual rules of shading and prevalence prescribed in the training ensemble, as well as (ii) certain \emph{implicit} contextual features that emerge as a result of the stochastic processes defined in the SCM. 

\begin{figure}[hbt]
\centering
\includegraphics[width=0.9\linewidth]{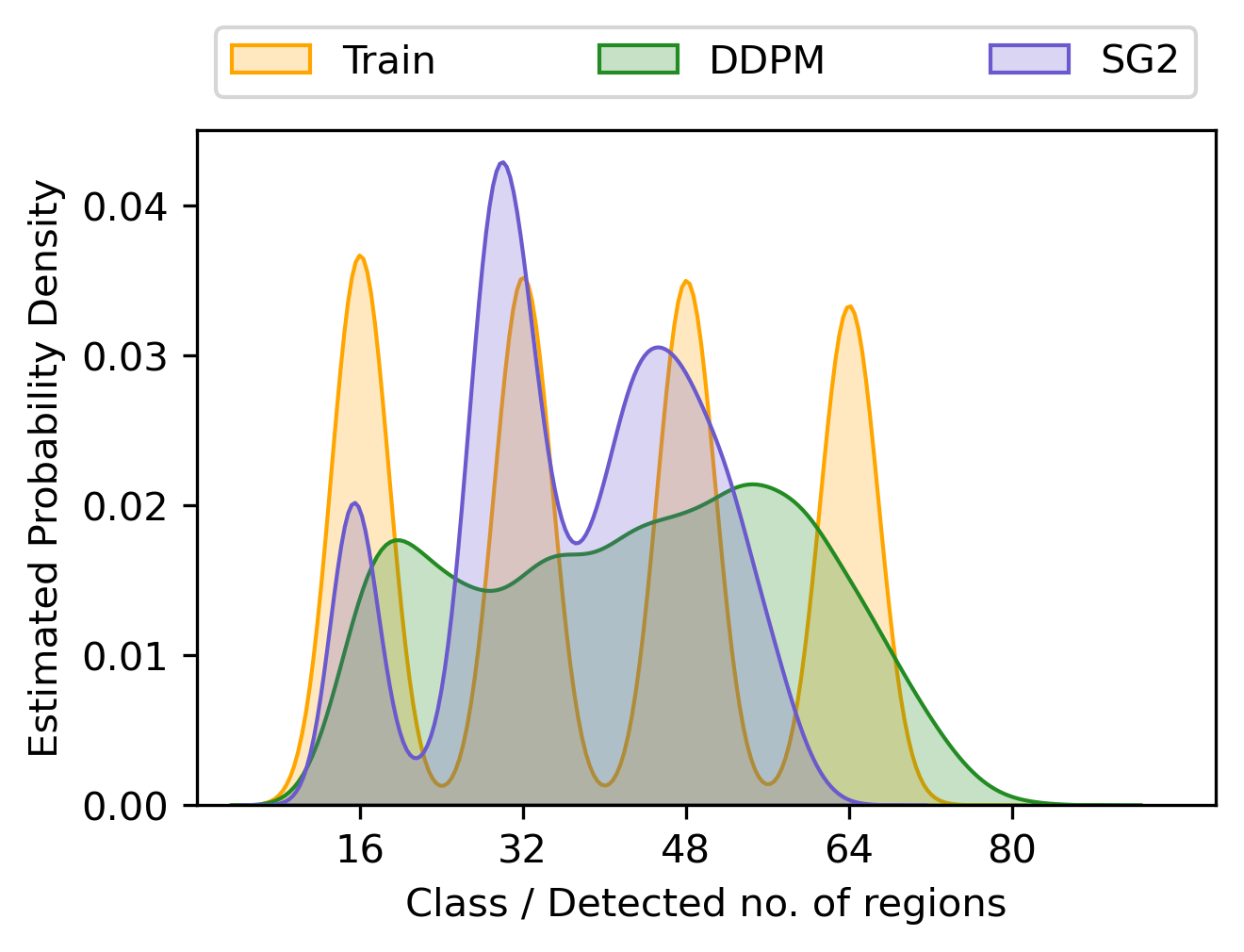} 

\caption{Class-prevalence results from the V-SCM. Both SG2 and DDPM fail to replicate the prescribed uniform class prevalence besides demonstrating interpolation between the four distinct classes in the training dataset. In addition, DDPM also extrapolates beyond the extreme class (class 64) generating realizations corresponding to class 80, which was absent in the training dataset.}
\label{voronoi_class_prevalence}
\end{figure}

The prescribed perfect correlation between grayscale intensity and area, within each image was observed to be lower in both DGM-generated ensembles. Approximately 13\% and 4\% of the SG2 and DDPM ensembles respectively demonstrated a Spearman rank-correlation $\rho<0.9$, indicating that the quantitative value of these realizations is partially lost. Next, per-image feature prevalence encoded as the number of Voronoi regions in an image was tested. Recall that, here, the number of regions defines class (see Sec. II.B.2 for class prediction on DGM-generated images). It was observed that neither DGM reproduced the prescribed uniform class prevalence (see Fig.~\ref{voronoi_class_prevalence}), although the DDPM demonstrated the better mode coverage.

\begin{figure}[h!bt]
\centering
\includegraphics[width=0.9\linewidth]{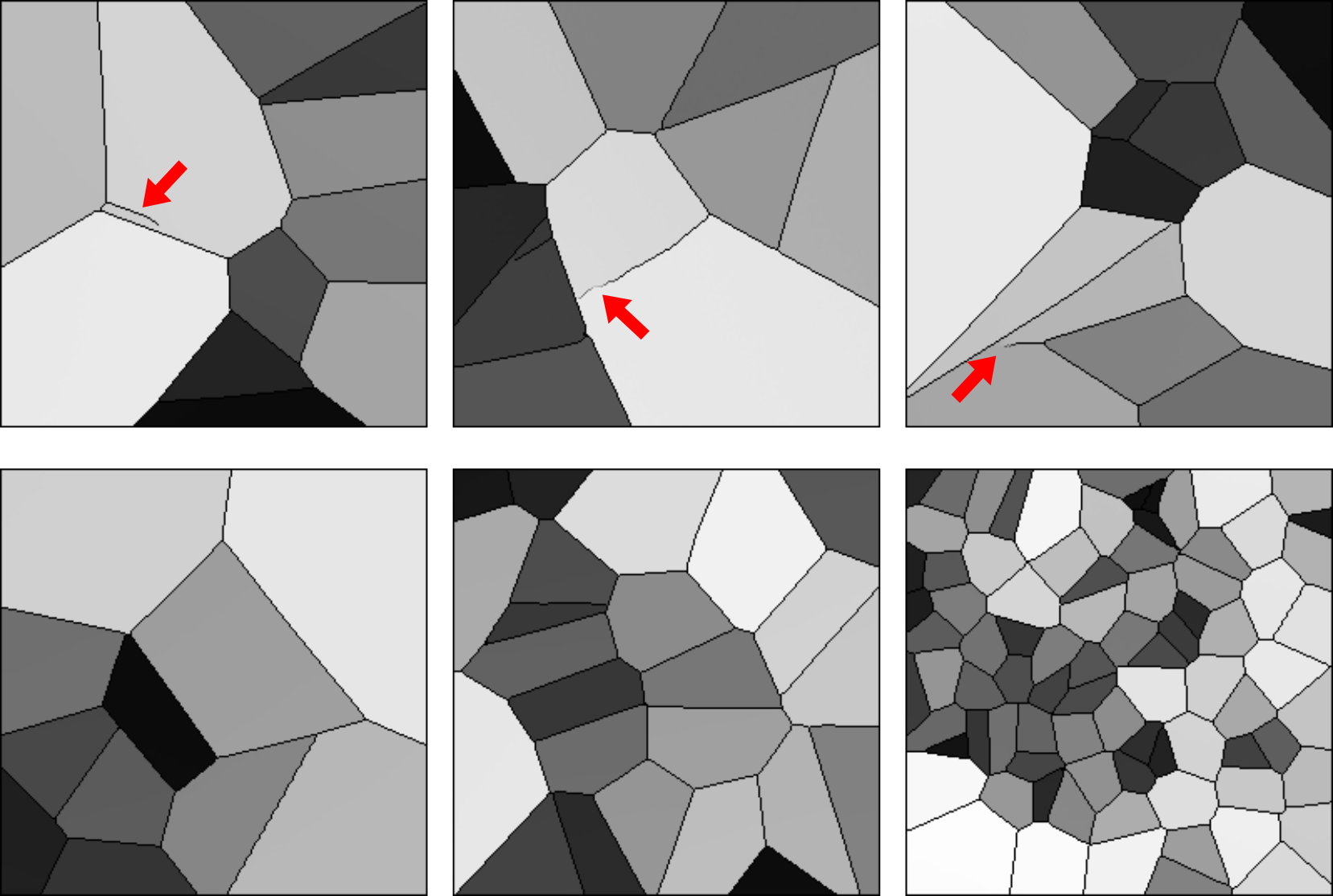} 

\caption{DDPM-generated samples from V-SCM exhibit implicit contextual errors like disjoint Voronoi edges (top row) and explicit errors like incorrect region count (bottom row). Although realizations in the bottom row are visually acceptable, the number of regions per-image indicating class is lower than (left), interpolated between (center), or extrapolated beyond, the classes in the training data (right)}
\label{SCM-V-bad_ones}
\end{figure}

Both DGMs were observed to interpolate between modes such that a substantial number of generated realizations are not any of the classes seen in the training data (see Fig.~\ref{voronoi_class_prevalence}). The extent of interpolation was \emph{less} for SG2, but still non-negligible. The SG2 model did not extrapolate beyond the extreme classes, but the DDPM unequivocally did (see Fig.\ref{SCM-V-bad_ones} bottom row). These observations could imply that interpolation and extrapolation are functionally equivalent within the assessed DGMs.

\begin{figure}[h!bt]
\centering
\includegraphics[width=0.9\linewidth]{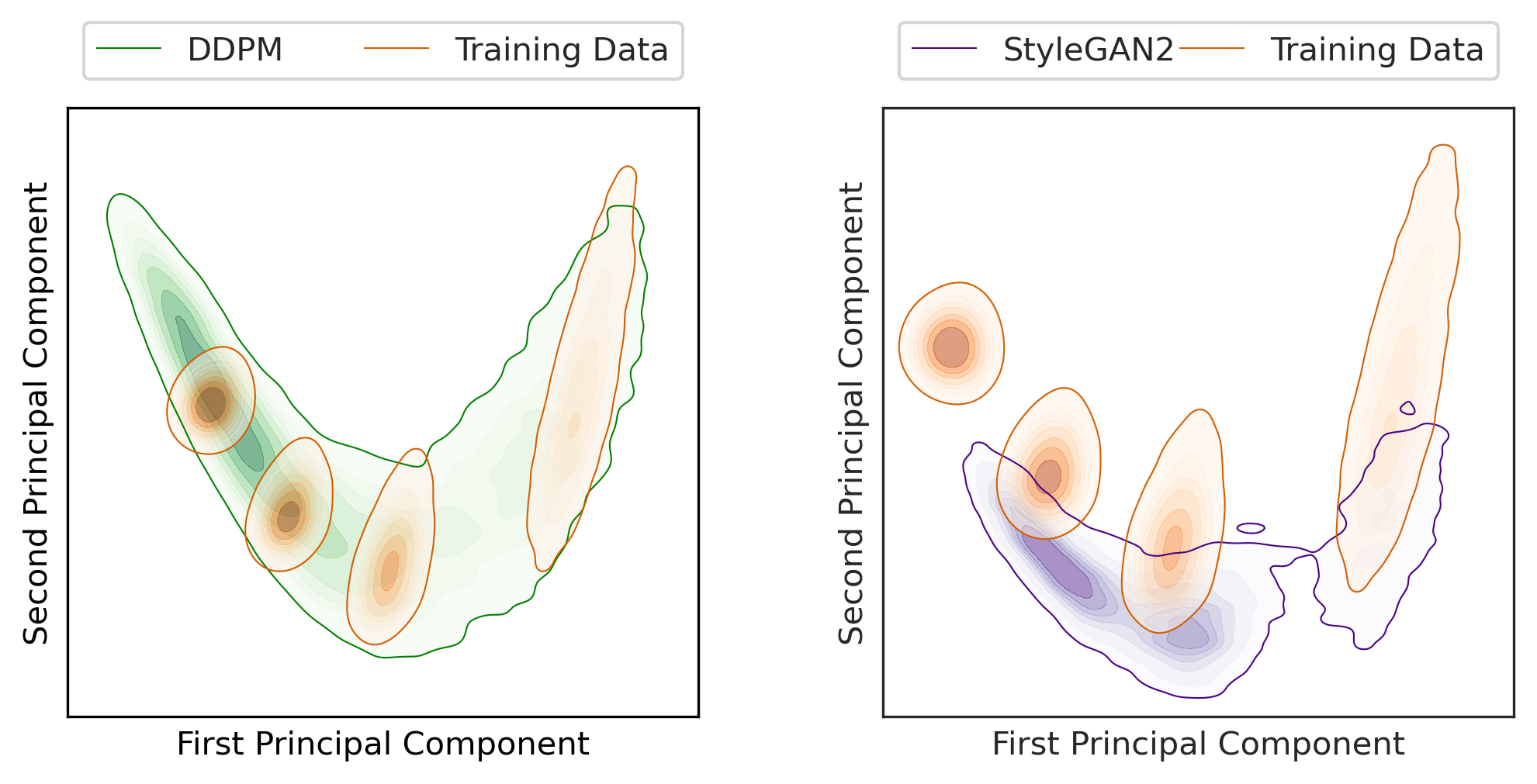} 

\caption{Results from the V-SCM. Principal component analysis of the statistics representing implicit context demonstrates that interpolation between classes also resulted in an interpolation of the emergent implicit context in case of DDPM (left), but not SG2 (right). The emergent implicit context was not exactly replicated by SG2 (right) as seen in the partial overlap between training and generated data.}
\label{voronoi_pca}
\end{figure}

Interpolation effects were further explored by assessing the implicit context typically arising in Voronoi diagrams \cite{boots2009spatial}. Given that Voronoi diagrams represent a unique solution to the space partitioning problem, variation in one statistic (e.g., number of regions) should affect all correlated statistics if the implicit context specific to Voronoi diagrams is exactly reproduced \cite{boots2009spatial}. The following per-image statistics (computed via a Python package \cite{nunez2018new}) were chosen to represent implicit context: number of Voronoi regions, number of junctions, junction density, mean and standard deviation of Voronoi edge lengths, mean and standard deviation of the area of a Voronoi region. Results from principal component analysis performed on these statistics are shown in Fig.~\ref{voronoi_pca}. It was observed that DDPM generated realizations followed the trend in implicit context defined by the training data. This result suggests that the DDPM generated a substantial number of realizations from new classes (via interpolation), but, perhaps more importantly, that those realizations may be genuine Voronoi diagrams. In Sec. V, this result is discussed further. On the other hand, this was not the case for SG2, which may be more prone to errors in implicit context for a similar interpolation between classes, suggesting that at least a fraction of the interpolated SG2 images may not be considered Voronoi diagrams. However, for the DDPM-generated ensemble, occasional errors in statistics representing implicit context were visually observed (see Fig.~\ref{SCM-V-bad_ones}), indicating that all implicit contextual features were not always perfectly reproduced. Thus, results from the V-SCM indicate that a large fraction of DDPM-generated images, but not all images, may be contextually correct in terms of quantitative meaning and class identity.

\subsection{Results from the F-SCM}
\begin{figure}[h!bt]
\centering
\includegraphics[width=0.9\linewidth]{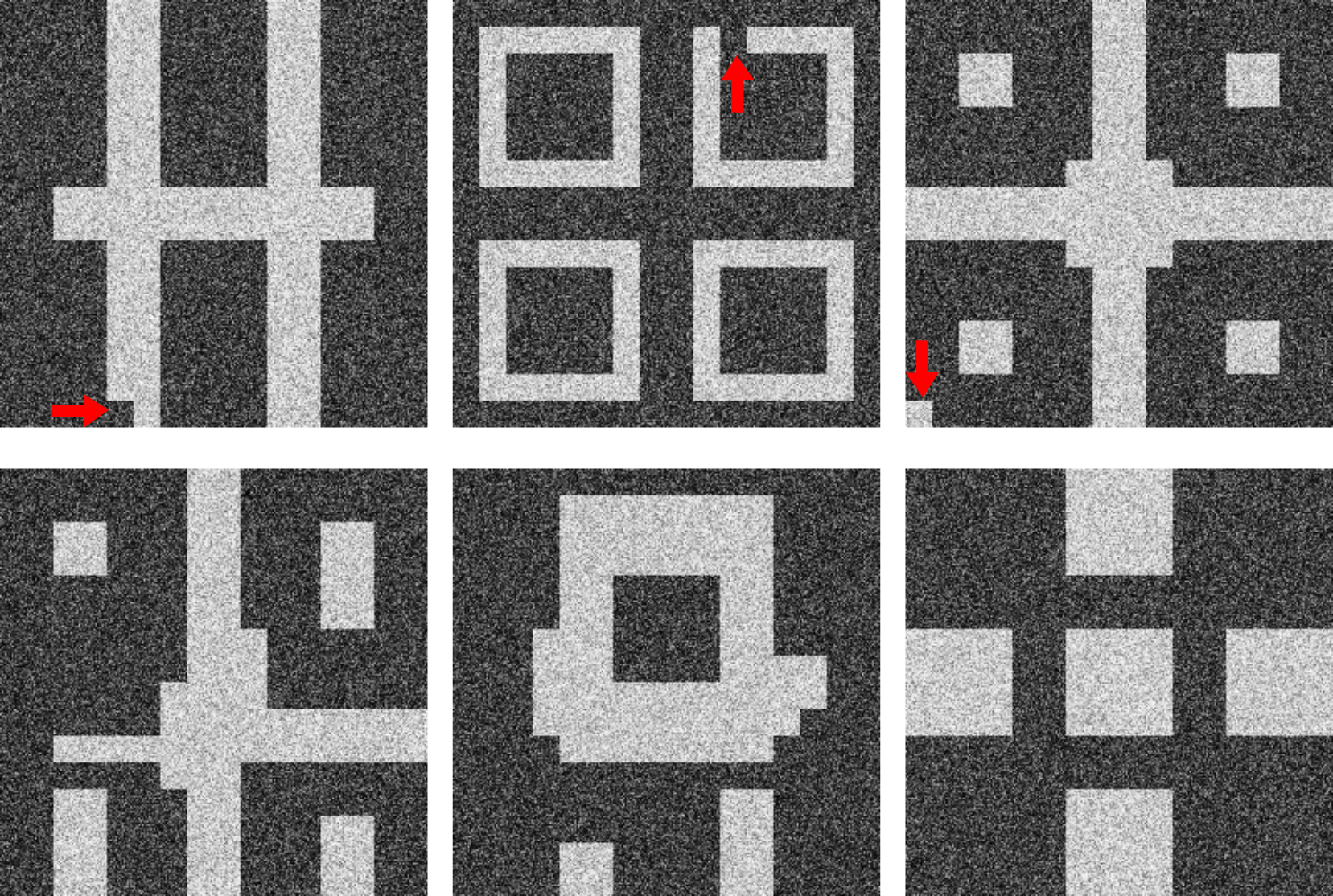} 
\caption{Contextually incorrect DDPM-generated samples from the F-SCM are shown. Top row: Minor errors in the foreground patterns due to a single misplaced tile were observed. Bottom row: Major errors in the class-specific foreground patterns were also observed. None of the foreground patterns in this row were present in the training data.}
\label{SCM-F-bad_ones}
\end{figure}

\begin{table}[htb]
\begin{center}
\caption{Results from the F-SCM. Percentage of acceptable realizations in an ensemble is reported for both DGMs and for four contextual constraints. DDPM slightly outperforms SG2 in most cases. Results for prevalence and position are reported together because these constraints jointly define the foreground structure representative of a class.}
\begin{tabular}{cccccc}
    \hline
    \multirow{2}{*}{Constraints}& \multirow{2}{*}{Measure of error} & \multicolumn{2}{c}{DDPM} & \multicolumn{2}{c}{SG2}  \\
    & & FG & BG & FG & BG\\
    \hline
    Prevalence & RMAE  & 99 & 99 & 98 & 98\\
     + Position & & & & & \\
    Intensity & $\chi^2$  & 0 & 0 & 0 & 9\\
    Texture & Moran's I  & 100 & 99 & 96 & 95\\

    \hline
\end{tabular}

\label{table_flags_results}
\end{center}
\end{table}

Quantitative results from the F-SCM that encodes joint contextual constraints in per-image feature prevalence, position, grayscale intensity, and texture are given in Table~\ref{table_flags_results}. The class-specific foreground patterns representing joint constraints in position and prevalence were correctly reproduced by both DGMs for over 98\% of the ensemble when tested via foreground template matching and relative mean absolute errors (RMAE). This is also visually evident in the DGM-generated images (see Fig.~\ref{SCM_good_ones}). 

However, errors in foreground patterns such as those shown in Fig.~\ref{SCM-F-bad_ones} were observed in about 1\% of the DDPM-generated ensemble. This is an important observation and the learning behavior of DDPM is discussed in detail in Section V. Note that the errors always occurred as misplaced or absent foreground tiles. Additionally, tiles which are never foreground in any class appeared as foreground in 0.1\% of the DDPM-generated ensemble, but never in the SG2-generated ensemble. This indicates that the DDPM learned individual motifs that create foreground patterns instead of entire image-level patterns.  
Texture arising from the randomness in pixel placement was correctly reproduced in over 95\% of SG2-generated ensembles and over 99\% of DDPM-generated ensembles as measured per tile via Moran's I \cite{moran1950notes}. Last, the prescribed per-image intensity distributions as measured via the $\chi^2$ goodness-of-fit test (at 95\% critical value) over each image were assessed. Except for a small fraction of SG-2 generated images, none of the images from either DGM were acceptable in terms of the foreground or background intensity distributions. These results might indicate some difficulty in learning multiple joint contextual constraints at once. Furthermore, the results also highlight the potential of a SCM-based evaluation approach, wherein contextual constraints are progressively added for the assessment of DGMs.

\subsection{Results from the VT-SOM}

\begin{figure}[hbt]
\centering
\includegraphics[width=0.9\linewidth]{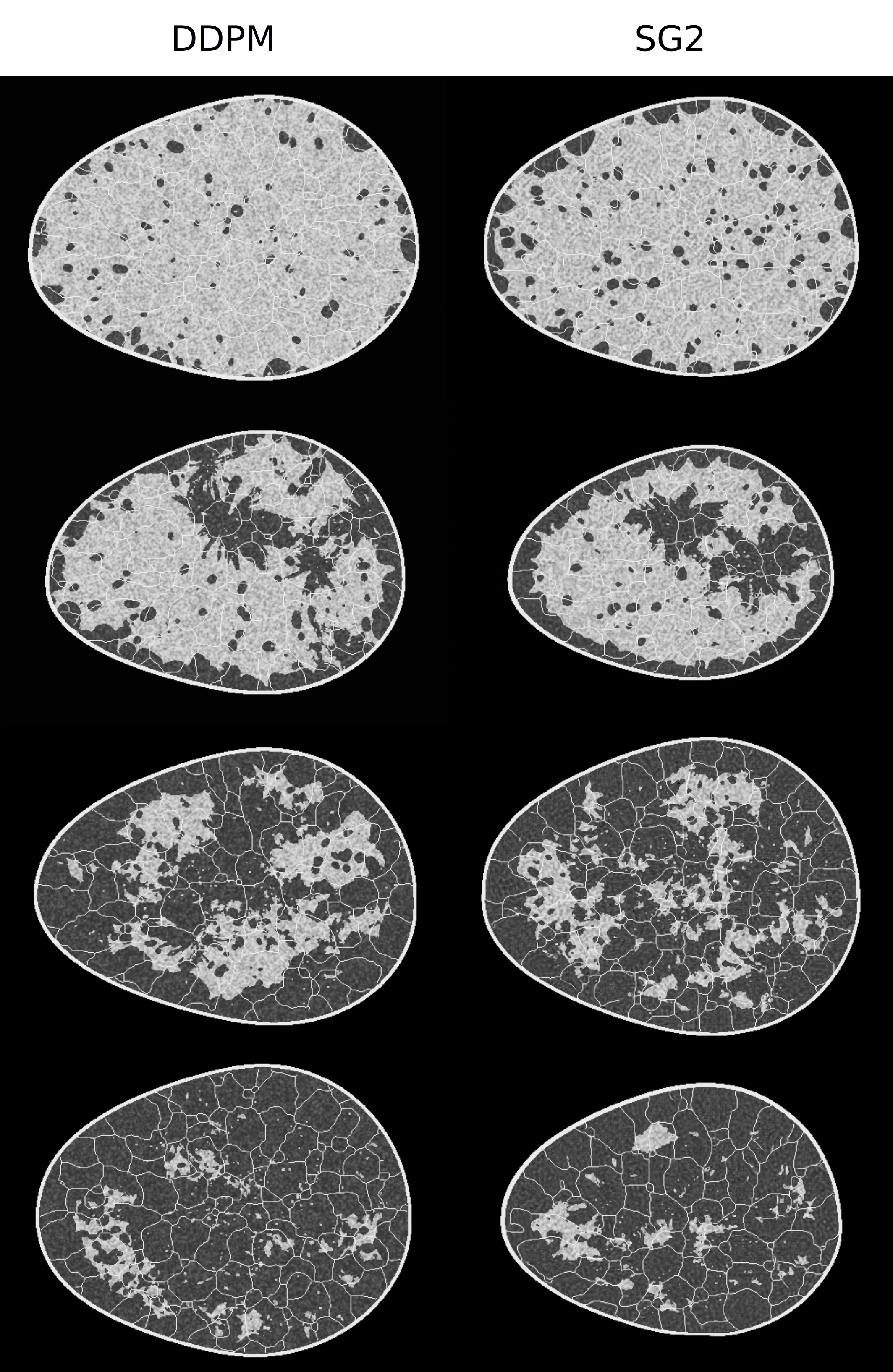} 

\caption{DGM-generated samples with high visual quality, corresponding to all four classes in the VT-SOM are shown. Recall that, here, the fat-to-glandular (F/G) ratio defines class.}
\label{victrei_res}
\end{figure}

Images generated from both DGMs demonstrated high visual similarity with the training data (see Fig.~\ref{victrei_res}) as well as low (\textless10) FID-10k scores; DDPM-generated images in particular, had distinctly superior visual image quality. Results from the VT-SOM demonstrate that DDPM clearly outperforms SG2 on most feature sets (see Table~\ref{table_victre_bestdgm}) included in the study, namely, texture features, morphology features, skeleton statistics, and the ratio of fatty to glandular tissue. (See Sec.III.B for a description of the evaluation framework.) 

\begin{figure}[hbt]
\centering
\includegraphics[width=0.9\linewidth]{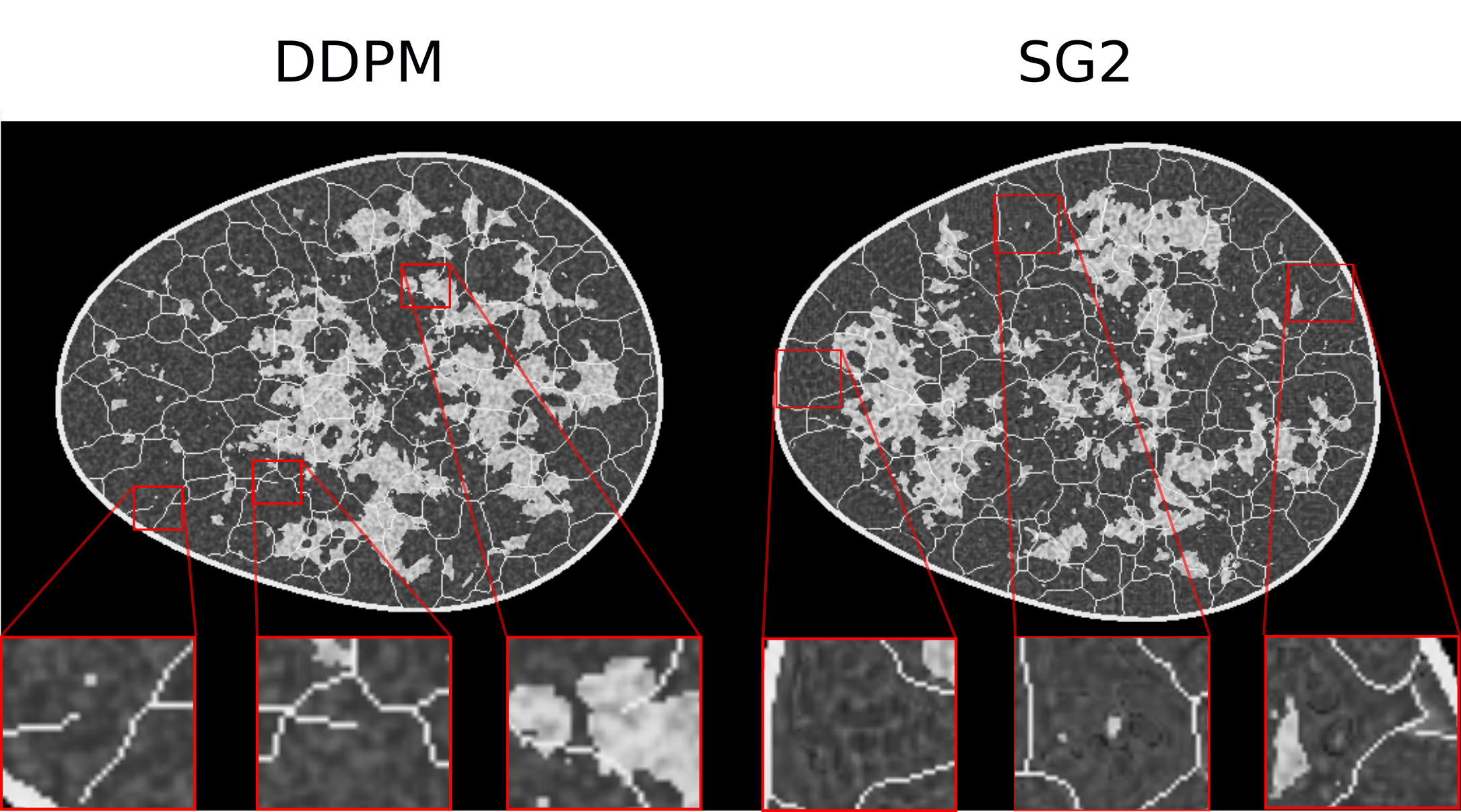} 
\caption{Samples from DGMs trained on VT-SOM show varied artifacts. DDPM images (left) exhibit strong visual quality but reveal structural errors like broken ligaments (inset). SG2 images (right) display artifacts in ligament structure, tissue texture, and shading (inset), not present in the training data.}
\label{victrei_bad}
\end{figure}

\begin{table}[h!tb]
\caption{Results from the VT-SOM for various feature families. Most feature families were better reproduced in the DDPM-generated ensemble as compared to the SG2-generated ensemble, as indicated by the lower KS statistic for the former. Analysis of class prevalence, coverage and density also demonstrate the superior performance of DDPM in representing all classes present in the training data.}
\begin{center}
\begin{tabular}{ccc}
    \hline
    Feature set / DGM & DDPM & SG2 \\
    \hline
    & \multicolumn{2}{c}{KS statistic} \\
    Texture features & \textbf{0.028} & 0.062 \\ 
    Morphology features & \textbf{0.049} & 0.28 \\
    Skeleton statistics & \textbf{0.006} & 0.19 \\
    F/G ratio & 0.23 & \textbf{0.11} \\
    Overall & \textbf{0.028} & 0.27\\
    \hdashline
    \\ 
    Class prevalence(\%) & 21,44,29,6 & 6,43,45,6 \\
    Class coverage \cite{naeem2020reliable} & 0.97, 0.96, 0.91, 0.91 & 0.35, 0.53, 0.74, 0.60  \\
    Class density \cite{naeem2020reliable} & 0.99, 1.01, 0.98, 1.02 & 1.02, 0.82, 0.99, 1.02  \\
    \hline
\end{tabular}

\label{table_victre_bestdgm}
\end{center}
\end{table}
This effect was particularly strong for morphology features (KS statistic values for DDPM: 0.049 and SG2: 0.278) and skeleton statistics (KS statistic values for DDPM: 0.006 and SG2: 0.195). However, SG2 outperformed DDPM in the reproducibility of F/G ratio. While SG2 replicated the four modes in the F/G ratio, DDPM demonstrated interpolation as well as substantial extrapolation, similar to results from the V-SCM. Some DDPM-samples with extreme F/G ratio, not seen in the training, were also observed; these samples were almost glandular tissue. Random samples of 200 images each from the training and DDPM-generated ensembles were visually inspected by non-experts for any immediately obvious errors. Occasional artifacts in ligament structures were visually noticeable in the DDPM-generated images (see Fig.~\ref{victrei_bad}). While major breaks in ligaments were observed in 1 in 6 images in the training ensemble, this rate doubled to 1 in 3 images in the DDPM-generated ensemble. These results indicate that even though DDPM outperformed SG2, it routinely synthesizes images with anatomical artifacts that can be spotted in the ligament structures even by a non-expert upon casual inspection. This should be taken into account before using the realizations for decision support.
Next, class-wise analysis was performed after predicting class labels on generated ensembles by employing a classifier with a VGG-16 \cite{simonyan2014very} backbone (see Sec. III.A). All four classes from the training ensemble were well represented in the DDPM-generated ensemble, but not in the SG2-generated ensemble. The class-wise prevalence percentages for the four classes were: 21\%, 44\%, 29\%, 6\% (DDPM), and 6\%, 43\%, 45\%, 6\% (SG2). The corresponding class-wise prevalence in the training data was: 10\%, 40\%, 40\%, 10\%. DDPM also demonstrated quantitatively superior or equivalent performance to SG2 in terms of class coverage and class density (see Table~\ref{table_victre_bestdgm}) for all classes. Thus, the class-wise results are in accordance with findings reported in literature that suggest that the DDPM largely alleviates the problem of mode collapse often observed in GANs \cite{xiao2021tackling}. 

\section{Discussion}
An evaluation of DDPM via stochastic models of context provides insights into the generative capacity of DDPM relevant to biomedical imaging, beyond conventional measures of image quality. Although contextual errors have been known to occur in the GAN family of DGMs \cite{deshpande2021method, dumont2021overcoming, muller2022diffusion, kelkar2023assessing}, to our knowledge, this is the first work to demonstrate and quantify various contextual errors in a diffusion-based generative model. Results from our studies demonstrate that impactful errors likely are present in every DGM-generated ensemble. The impact of those errors is task-dependent and therefore should be studied case-by-case. 

The relevance of the evaluation approach to medical imaging applications employed in this work lies not in anatomical realism, but in the logically analogous representation of contextual attributes relevant to biomedical imaging. For example, one work \cite{dumont2021overcoming} reported contextual errors in GAN-generated images such as misplaced pacemakers in chest radiographs. Analogously, we observed misplaced tiles in DDPM-generated Flags-SCM images, thus exposing the capacity of DDPMs to misplace features in forbidden areas. Other examples of biomedical imaging scenarios that involve the studied contextual attributes include: (i) pathology images, wherein the cell-specific size, intensity distribution and per-image prevalence may be characteristic of different pathologies and (ii) the relative positions of organs, and the per-image prevalence of ribs in a chest radiograph. The two examples are respectively analogous to: (i) the Voronoi SCM, which encodes context via shading and prevalence at multiple length scales, and (ii) the Alphabet SCM, which encodes context via per-image prevalence and relative positions of letters.

\begin{figure}[h!bt]
\centering
\includegraphics[width=0.9\linewidth]{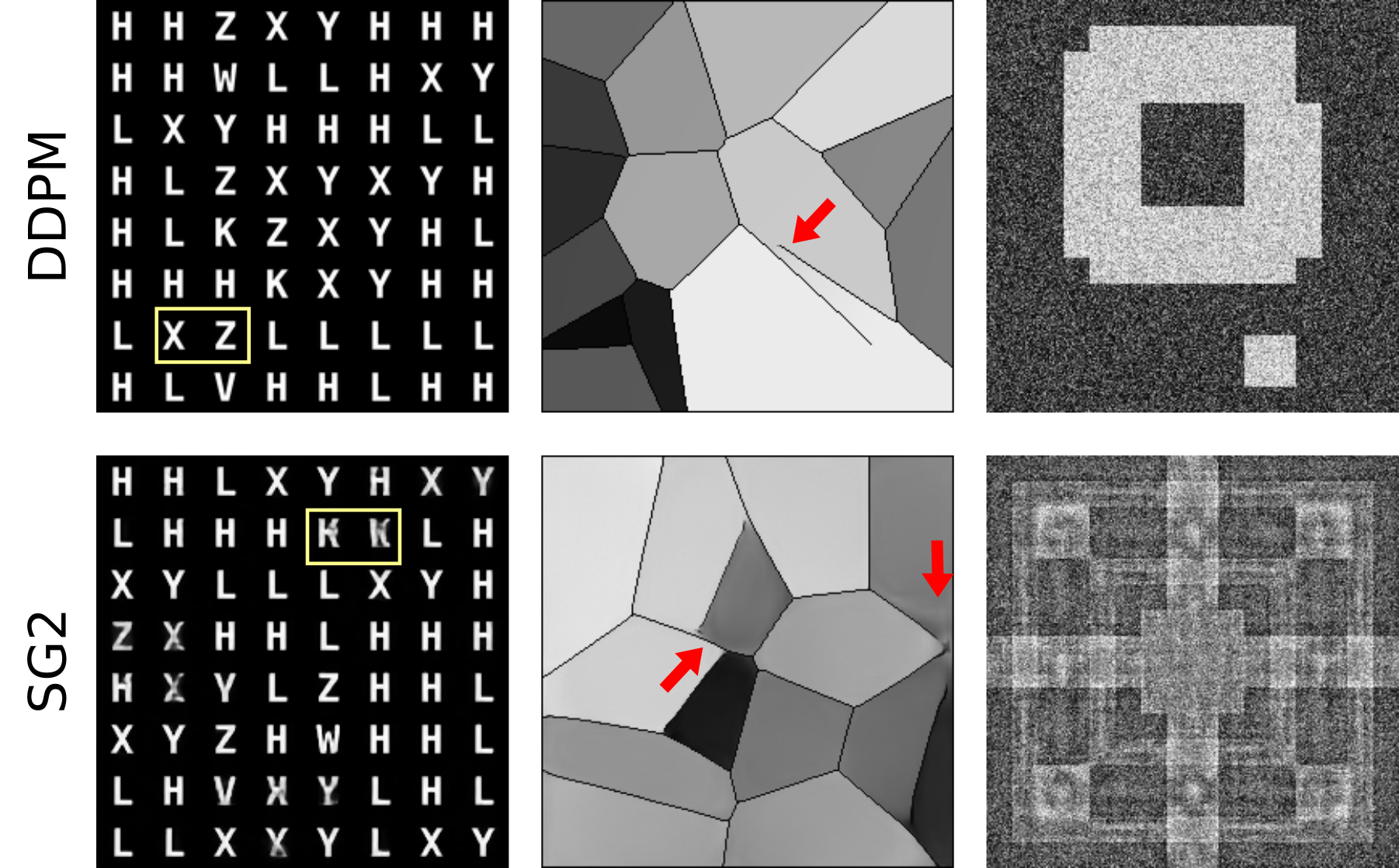} 
\caption{Examples of artifacts present in generated realizations from DDPM and SG2 are shown for the three SCMs. Errors in DDPM-generated images demonstrate misplaced but distinct motifs from the training data, whereas errors in SG2-generated images demonstrate malformations or blending of distinct motifs. This effect is seen across all SCMs.}
\label{SCM-ddpm-vs-sg2}
\end{figure}

One key finding is that implicit context \emph{for new classes} was very well reproduced in the DDPM-generated Voronoi SCM ensemble (see Fig.~\ref{voronoi_pca}). This is particularly important because the implicit context that defines a genuine realization is known for Voronoi diagrams \cite{boots2009spatial}. Thus, the ``correctness'' of a V-SCM realization actually can be measured. The corollary is that new experiments can be designed to determine the amount and type of training data required to generate $N$ images of a desired quality, as is necessary to validate data augmentation schemes. This is in stark contrast to traditional measures of DGM-generated image quality which do not necessarily indicate at all the quality of image \emph{content}. For example, there is no known set of statistics or formulas that defines what a human heart ought to look like; and the extent to which natural human perception impacts quality assessment of images is unknown, and may well be unquantifiable \cite{emrith2010measuring}. Thus, SCMs provide a domain-apt ground truth where none otherwise is known. The particular SCMs employed in the present work were chosen for ease of inspection and discussion. The key property that a designed SCM should have is post-hoc recoverability of context. This can be achieved in numerous ways but perhaps the simplest is to selectively constrain an existing stochastic model of interest. This is what was done to the VICTRE model. That model already was demonstrated to be anatomically adequate for some virtual imaging trials \cite{badano2018evaluation}. By imposing an exact distribution on intensity, a known correlation length on the arrangement of those, and a slightly processed ligament structure, several recoverable contexts were created without losing the overall anatomical realism. 

A second result, also pertaining to interpolation across classes, was observed via the F-SCM. Specifically, interpolated instances in the DDPM-generated ensemble were not merely a linear combination of foregrounds in the training data. Furthermore, some instances also violated the regions forbidden as foreground across all classes. However, the grid in the F-SCM design, or size of a single tile, was correctly learned by the DDPM, and it seems that this knowledge was employed in class interpolation. Thus, although the DDPM correctly identified the relevant local scale in the formation of patterns, it did not perfectly capture the image-level context in the F-SCM. On a more complex dataset: the VT-SOM, the DDPM largely failed to capture all contextual constraints at once. Despite the high visual quality of DDPM images, major errors in ligament formation were identified even by a non-expert in about 30\% of the ensemble. Furthermore, unrealistic extrapolation beyond all classes in the training data was observed in the V-SCM and F-SCM, potentially due to the likelihood-based approach of DDPM that also contributes to excellent mode coverage \cite{xiao2021tackling}. It is possible that a class-conditioned DDPM may partially alleviate this issue, but such an evaluation is beyond the scope of this work.

\begin{figure}[h!bt]
\centering
\includegraphics[width=0.9\linewidth]{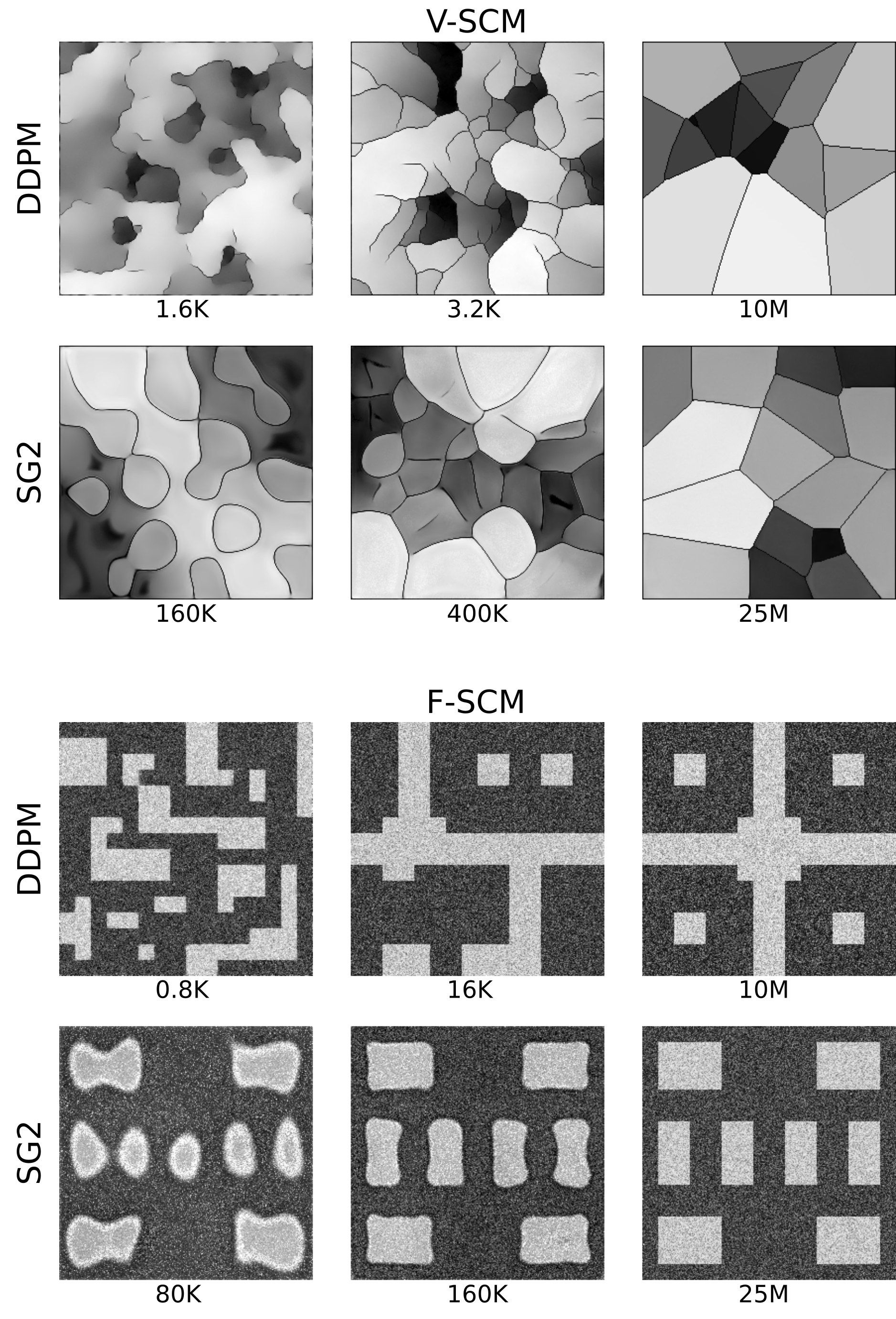} 

\caption{Visually interesting but random examples from the training trajectories of DDPM and SG2 models employed on V-SCM and F-SCM datasets are shown. The training step corresponding to each image is indicated below the image, and represents the number of images seen in training. DDPM seemed to first learn local elements that constitute the expected structure, while SG2 seemed to learn image structure by moulding blob-like elements.}
\label{training_trajectories}
\end{figure}

Even these simple SCMs reveal stark differences in image representation within the two DGM paradigms: DDPM and StyleGAN2, particularly, via the nature of the artifacts (see Fig.~\ref{SCM-ddpm-vs-sg2}) and training trajectories Fig.~\ref{training_trajectories}. While artifacts in DDPM generally seemed to involve misplaced but correct motifs, the artifacts in SG2 demonstrated a blending of various motifs within the same image. Similar differences were observed in the training trajectories (Fig.~\ref{training_trajectories}). DDPM seemed to first learn local elements required to construct image-level structure followed by combinations of these elements, while SG2 seemed to learn image structure through blob-like elements. 

Each SCM in this work represents a different context; together, the SCMs constitute a readily interpretable and intuitive method for the objective assessment of DGMs. The SCMs successively encode an increasing number of contextual constraints; this enables a step-wise evaluation of the capacity of a DGM to reproduce individual and joint contextual constraints (see Table~\ref{table_scms}). For example, DDPM almost perfectly replicates the letter prevalences in the A-SCM, and largely reproduces the contextual constraints of shading and prevalence in the V-SCM, but fails to reproduce the intensity distributions in the F-SCM. This suggests that the joint replication of multiple contextual constraints at once remains a challenge for DDPM. 

\section{Conclusion}
In this work, an instance of the denoising diffusion probabilistic model (DDPM) was evaluated to gain insights into its capacity to reproduce contextual attributes analogous to anatomical constraints present in medical imaging scenarios. The DDPM-generated ensembles in this study demonstrated low contextual error-rates, but none of the ensembles reproduced the expected context perfectly.
This evaluation goes beyond earlier evaluations of DDPM that employed ensemble-based evaluation measures designed for natural images, or conventional measures of image quality. We anticipate that the employed evaluation framework might yield insights into emerging DGMs and have a broader impact on decision-making and DGM benchmarking.

\bibliographystyle{IEEETran} 
\bibliography{IEEEabrv,references}

\end{document}